\documentclass[12pt]{amsart}
\usepackage{euscript}
\usepackage{multicol}
\usepackage{amssymb}
\usepackage{amsmath}
\usepackage{times} 

\newtheorem{theorem}{Theorem}[section]
 
\newtheorem{lemma}{Lemma}[section]
\numberwithin{equation}{section}
\begin{document}
\title[Follytons and the Removal of Eigenvalues]
{Follytons\\ 
and the Removal of Eigenvalues\\ 
for Fourth Order Differential Operators}
\author[Hoppe, Laptev and \"Ostensson]{J. Hoppe, A. Laptev and J. \"Ostensson}
\begin{abstract}
A non-linear functional $Q[u,v]$ is given that 
governs the loss, respectively gain, of (doubly degenerate) eigenvalues
of fourth order differential operators 
$L = \partial^4 + \partial\,u\,\partial + v$ on the line. Apart from 
factorizing $L$ as $A^{*}A + E_{0}$, providing several explicit 
examples, and deriving various relations between $u$, $v$ and 
eigenfunctions of $L$, we find $u$ and $v$ such that $L$ is isospectral to 
the free operator $L_{0} = \partial^{4}$ up to one 
(multiplicity 2) eigenvalue $E_{0} < 0$. Not unexpectedly, this choice
of $u$, $v$ leads to exact solutions of the 
corresponding time-dependent PDE's. 
\end{abstract}

\email{hoppe@math.kth.se, laptev@math.kth.se, jorgen@math.kth.se}
\subjclass{Primary 34L40; Secondary 34L30.}
\maketitle

\section{Factorization of the operator 
$L = \partial^4 + \partial\,u\,\partial + v$.}

Let us assume that $u$ and $v$ are real-valued functions and 
$u, v\in \mathcal{S}\left(\mathbb{R}\right)$, 
where $\mathcal{S}\left(\mathbb{R}\right)$
denotes the Schwarz class of rapidly decaying functions. Let $L$
be a linear fourth order selfadjoint operator in $L^2\left(\mathbb{R}\right)$
\begin{equation}\label{L} 
L := \partial^4 + \partial\,u\,\partial + v
\end{equation}
defined on functions from the Sobolev class $H^4(\mathbb{R})$.
This operator is bounded from below and we assume that 
its lowest eigenvalue $E_0<0$ is of double multiplicity
and therefore there exist two orthogonal in $L^2(\mathbb{R})$ eigenfunctions
$\psi_+$ and $\psi_-$ satisfying the equation
\begin{equation}\label{groundstate}
L\,\psi = E_{0}\,\psi.
\end{equation}
As shown in the appendix, the Wronskian 
\begin{equation}
\label{wronsk}
W(x):= \psi_{+}(x)\,\psi_{-}'(x) -  \psi_{-}(x)\,\psi_{+}'(x) 
\end{equation}
is necessarily non-vanishing, $W(x)\not= 0$, $x \in \mathbb{R}$.
Let us try to factorize $L-E_{0}$ as 

\begin{equation}   
\label{factorization}
A^{*} A = \left(- \partial^2 - f \partial + g - f'\right) 
\left(- \partial^2 + f \partial + g\right), 
\end{equation}
with $f$ and $g$ real-valued. Clearly, 
\begin{equation}
  \label{odefg}
\left\{
\begin{array}{ll}
f' + f^2 + 2g &= - u\\
g^2 - (fg + g')' &= v - E_{0}.
\end{array}
\right.
\end{equation}
Instead of discussing these non-linear differential equations
directly, let us express $f, g, u$ and $v$ in terms of the 
functions $\psi_{+}$, $\psi_{-}$. Straightforwardly, one finds that
since $\psi_{+}$ and $\psi_{-}$ are eigenfunctions of $A^{*} A$ with 
eigenvalue $0$, we have $A \psi_{+} = A \psi_{-} = 0$, which implies 
\begin{equation}
  \label{eqnfg}
\left\{
\begin{array}{rl}
f\,W &= W'\\
-g\,W &= \psi_{+}'\,\psi_{-}'' - \psi_{+}''\,\psi_{-}' =: W_{12},   
\end{array}
\right.
\end{equation}
while
$\left(L - E_{0}\right) \psi_{+} = \left(L - E_{0}\right) \psi_{-} =
0$ implies 
\begin{equation}
  \label{eqnuv}
\left\{
\begin{array}{rl}
u\,W &= 2 \, W_{12} - W'' + \epsilon\\
\left(v - E_{0}\right)\,W&= u\,W_{12} + W_{12}'' -W_{23},
\end{array}
\right.
\end{equation}
where $\epsilon$ is an integration constant and
\begin{equation}
  \label{W23}
W_{23} := \psi_{+}''\,\psi_{-}''' - \psi_{+}'''\,\psi_{-}''
\end{equation}
is expressible in terms of $W$ and $W_{12}$ via
\begin{equation}
  \label{W23relation}
W\,W_{23} = W_{12}'\,W'-W_{12}\,W'' + W_{12}^2.
\end{equation}
Equations \eqref{eqnfg} say that 
\begin{equation}
  \label{fg}
f =  \frac{W'}{W}, \quad g = - \frac{W_{12}}{W}. 
\end{equation}
Since $u\,W + W'' - 2\,W_{12}$ vanishes at infinity,
$\epsilon$ has to be $0$, and one finds, using equations 
\eqref{eqnuv}-\eqref{W23relation}, that

\begin{alignat}{4}
  \label{eqnu}
u &= \frac{2\,W_{12} - W''}{W} \\
\label{eqnv}
v - E_{0} &= \frac{W_{12}^2}{W^2} + 
\left(\frac{W_{12}'}{W}\right)'.
\end{alignat}
Note that 
\begin{equation}
  \label{darbouxL}
\tilde{L} := A A^{*} + E_{0} = L + 4\,\partial\,f'\,\partial + 2\,f\,g' 
-f\,f''+f''' 
\end{equation}
will be isospectral to $L$, apart from $E_{0}$, which has been
removed. To see why $E_{0}$ is not an eigenvalue of 
$\tilde{L}$, let us for simplicity assume that 
$u, v \in C_{0}^{\infty}\left(\mathbb{R}\right)$, say that supp$\,u$,
supp$\,v \subset (-c,c)$. Then, 
\begin{equation*}
\begin{array}{rl}
\psi_{+}(x) = \alpha_{1}\,e^{-\kappa x}\,\cos\,(\kappa x) + \beta_{1}\, 
e^{-\kappa x}\,\sin\,(\kappa x)\\
\psi_{-}(x) = \alpha_{2}\,e^{-\kappa x}\,\cos\,(\kappa x) + \beta_{2}\, 
e^{-\kappa x}\,\sin\,(\kappa x)
\end{array},
\quad x > c,
\end{equation*}
where $E_{0} = - 4\,\kappa^4$, $k>0$. This implies
\begin{equation*}
\begin{array}{rl} 
W(x) &= \kappa\,e^{-2\kappa x}\,\left(\alpha_{1}\,\beta_{2} -
\beta_{1}\,\alpha_{2}\right)\\
W_{12}(x) &= 2\,\kappa^3\,e^{-2\kappa x}\,\left(\alpha_{1}\,\beta_{2} -
\beta_{1}\,\alpha_{2}\right)
\end{array},
\quad x > c.
\end{equation*}
(note that the bracket does not vanish, since $\psi_{+}$ and $\psi_{-}$
are linearly independent.) This (and a similar investigation at the
other end) implies that 
\begin{equation*}
f(x) = \mp 2\,\kappa, \quad g(x) = -2\,\kappa^2, \quad \mbox{for } \pm\,x >c.
\end{equation*}
Since $\tilde{L}\psi = E_{0}\,\psi$ implies $A^{*} \psi = 0$, we
obtain
\begin{equation*}
\psi'' - 2\kappa\,\psi' + 2\kappa^2 \psi = 0, \quad x > c.
\end{equation*} 
It clearly follows that $\psi$ cannot be in
$L^{2}\left(\mathbb{R}\right)$ unless it vanishes identically.

\smallskip
Before giving some explicit examples, let us make some comments
concerning the problem of actually finding $f$ and $g$, or $\psi_{+}$ and
$\psi_{-}$, when $u$ and $v$ are given. Instead of 
solving the non-linear system \eqref{odefg}, or the spectral problem
\eqref{groundstate},  one may also try to solve the Hirota-type
equation which follows from \eqref{eqnu}, \eqref{eqnv}
\begin{equation}
  \label{hirota}
  4\,\left(v-E_{0}\right) = \left(\frac{W''}{W} + u\right)^2 
+ 2\,\left(\frac{W'''}{W} + u' + u\,\frac{W'}{W}\right)',
\end{equation}
and which for $u \equiv 0$ reads
\begin{equation*}
  \label{hirota2}
4\,\left(v-E_{0}\right)\,W^{2} = 
2\,\left(W''''\,W - W'''\,W'\right) + W''^{2}.
\end{equation*}
Once $W\,(\neq 0)$ is obtained, $f$ and $g$ can be given by the equations 
\eqref{fg}. With $f$ and $g$ defined in this way \cite{Hoppe1},
equation \eqref{odefg} is satisfied and the factorization 
\eqref{factorization} is valid.

\smallskip
Note also the following: the functions $\psi_{+}$ and $\psi_{-}$ are
solutions of $A\,\psi = 0$, i.e. 
\begin{equation*}
-\psi'' + f\,\psi' + g\,\psi = 0.
\end{equation*}
By writing  
\begin{equation*}
\psi_{\pm} = \sqrt{W}\,\phi_{\pm} ,
\end{equation*}
one finds that $\phi_{+}\,\phi_{-}' - \phi_{+}'\,\phi_{-} =1$
and that $\phi_{\pm}$ are (oscillating) solutions of the
equation in Liouville form 
\begin{equation*}
-\phi'' + \left(g +\frac{3}{4}\left(\frac{W'}{W}\right)^2 -
 \frac{1}{2}\,\frac{W''}{W}\right)\,\phi = 0,
\end{equation*}
i.e. associated to a second order self-adjoint diffential operator.
\section{Addition and removal of eigenvalues.}
Although adding and removing eigenvalues may be thought to be a 
procedure that can be read both ways (symmetrically), the steps
involved are actually quite different in both cases (in particular, 
it is not yet clear, which conditions on $u$ and $v$ allow for the addition
of a doubly degenerate eigenvalue below the spectrum of $\partial^4 +
\partial\,u\partial + v$).
Let us therefore 'summarize' them separately, in both cases starting
from a given operator 
\begin{equation*}
L_{n} := \partial^4 + \partial\,u_{n}\partial + v_{n},\quad n \in \mathbb{N}, 
\end{equation*}  
and the equation \eqref{hirota} with $u, v$ replaced by $u_{n},
v_{n}$. This equation shall be referred to as \eqref{hirota}$_{n}$.\\
\textbf{Removal of eigenvalues:}\\
\textit{1}. Solve \eqref{hirota}$_{n}$ 
$\left(\mbox{with } E_{0} \rightarrow E_{0}^{(n)} = -4 \kappa_{n}^4\right)$ 
for $W_{n} := W$ $\left( \rightarrow 0 \mbox{ at infinity}\right)$ and 
define $W_{12}^{(n)}$ as $\frac{1}{2}\left(W_{n}\,u_{n} +
W_{n}''\right)$, as is natural in accordance with equation
\eqref{eqnu}$_{n}$. Alternatively, if $\psi_{\pm}^{(n)}$ are known,
calculate $W_{n}$ and $W_{12}^{(n)}$ via their definitions, i.e. as 
\begin{equation*}
\begin{array}{rl}
W_{n} &= \psi_{+}^{(n)}\,{\psi_{-}^{(n)}}' - \psi_{-}^{(n)}\,{\psi_{+}^{(n)}}' \\
W_{12}^{(n)} & = {\psi_{+}^{(n)}}'\,{\psi_{-}^{(n)}}'' - 
{\psi_{+}^{(n)}}''\,{\psi_{-}^{(n)}}' .
\end{array}
\end{equation*}
\textit{2}. Define $f_{n}$ and $g_{n}$ according to \eqref{fg}$_{n}$,
thus solving the system \eqref{odefg}, and obtaining the factorization 
\begin{equation*}
L_{n} =A_{n}^{*} A_{n} - 4\kappa_{n}^4.
\end{equation*}
\textit{3}. The operator
\begin{equation*}
\tilde{L}_{n} =A_{n} A_{n}^{*} - 4\kappa_{n}^4 =: L_{n-1}
\end{equation*}
will then be isospectral to $L_{n}$ apart form the lowest eigenvalue 
$E_{0}^{(n)} = - 4\kappa_{n}^4$ (of multiplicity 2), which has been
removed.\\
\textbf{Addition of eigenvalues:}\\
\textit{1}. Solve \eqref{hirota}$_{n}$ 
$\left(\mbox{with } E_{0} \rightarrow E_{0}^{(n+1)} = -4 \kappa_{n+1}^4\right)$ 
for $\hat{W}_{n+1} := W \sim e^{\pm 2 \kappa_{n+1} x}$, as $x
\rightarrow \pm \infty$, i.e. $\hat{W}_{n+1}$ diverging at 
infinity and non-vanishing for finite $x$. (As mentioned above,
conditions on $u_{n}$, $v_{n}$ ensuring the existence of
$\hat{W}_{n+1}$ are still unclear.) \\
\textit{2}. Define $W_{n+1} := \frac{1}{\hat{W}_{n+1}}$, which will 
then solve the (more complicated looking) equation
\begin{alignat}{4}
  \label{addeqn1}
&40\,\frac{W'^{4}}{W^{4}}-2\,\frac{W''''}{W}+14\,\frac{W'''\,W'}{W^{2}}+
13\,\frac{W''^{2}}{W^{2}}-64\,\frac{W''\,W'^{2}}{W^{3}}+\\
\notag
&2u''+ u^2-2u'\,\frac{W'}{W}+2u\left(\frac{W'^{2}}{W^{2}}-
2\left(\frac{W'}{W}\right)'\right) = 16 \kappa^4 + 4 v
\end{alignat} 
(with $u$, $v$ $\rightarrow$ $u_{n}$, $v_{n}$ and $\kappa \rightarrow
\kappa_{n+1}$). 
In fact, \eqref{addeqn1} is equivalent to 
\begin{alignat*}{4}
&-2f'''+6ff''+7f'^2-8f'f^2+f^4+2u\left(f^2-2f'\right)-2u'f+u^2+2u'' \\
\notag
&= 4v+16\kappa^4
\end{alignat*}
$\left(\mbox{via } f = \frac{W_{n+1}'}{W_{n+1}} =:f_{n+1},\,u, v \rightarrow
u_{n}, v_{n} \mbox{ and } \kappa \rightarrow \kappa_{n+1}\right)$ that 
arises in the factorization of $L_{n+1}$. \\
\textit{3}. Write
\begin{equation*}
L_{n} = A_{n+1} A_{n+1}^{*} - 4\kappa_{n+1}^4
\end{equation*}
(implying $2\,g_{n+1} := 3\,f_{n+1}' - f_{n+1}^2 - u_{n}$).
\\
\textit{4}. Then, 
\begin{equation*}
L_{n+1} := A_{n+1}^{*} A_{n+1} - 4\kappa_{n+1}^4,
\end{equation*}
will be isospectral to $L_{n}$ apart from having one additional
(doubly degenerate) eigenvalue $E_{0}^{(n+1)}$ below the spectrum of
$L_{n}$.

\section{A non-linear functional $Q$ and a system of PDE's
associated with the operator $L$.}
 
As observed 100 years ago \cite{Sch}, the operator $L = \partial^4 + 
\partial\,u\,\partial + v$ has a unique 4'th root in the form 
$L^{1/4} := \partial + \sum_{k=1}^{\infty} l_{k}(x) \partial^{-k}$. 
%
%
Define $M$ to be the positive (differential operator) part of any 
integer power of $L^{1/4}$. Then it is well known, that 
\begin{equation*}
L_{t} = \left[L, M\right],
\end{equation*}
where $L_{t}$ is the operator defined by
$
L_{t}\,\varphi = \partial\,u_{t}\,\partial\,\varphi + v_{t}\,\varphi, 
$
consistently defines evolution equations 
(for $u = u(x, t)$, $v = v(x, t)$) that have infinitely many conserved 
quantities (i.e. functionals of $u$ and $v$, and their spatial derivatives, 
that do not depend on $t$). We shall make use of this by letting  
\begin{equation*}
M := 8\left(L^{3/4}\right)_{+} = 8\,\partial^3 + 6\,u\,\partial + 3\,u', 
\end{equation*}
and focusing on the quantity
\begin{equation}
  \label{functionalQ}
Q[u, v] := \int_{\mathbb{R}} \left(48\,v^2 +
\frac{5}{4}\,u^4 - 12\,u^2\,v - 40\,u\,v'' - 13\,u\,u'^2 +
9\,u''^2\right) \,dx. 
\end{equation}
This quantity does not change when $u$ and $v$ evolve
according to
\begin{equation}
  \label{KdVtype}
\left\{
\begin{array}{rl}
u_{t} &= 10\,u''' + 6\,u\,u' - 24\,v' \\
v_{t} & = 3\left(u''''' + u\,u''' + u'\,u''\right) - 8\,v''' - 6\,u\,v'.
\end{array}
\right.
\end{equation}
Formula \eqref{darbouxL} for $\tilde{L} = \partial^{4}
+\partial\,\tilde{u}\,\partial + \tilde{v}$ implies that 
\begin{equation}
  \label{darbouxL2}
\left\{
\begin{array}{rl}
\tilde{u} - u &= 4\,f'\\
\tilde{v} - v &= 2\,f\,g' -f\,f''+ f'''.
\end{array}
\right.
\end{equation}
By using the asymptotic properties of $f$ and $g$ ($f \rightarrow \mp 
2\,\kappa$, $g \rightarrow -2\kappa^{2}$, as $x \rightarrow \pm
\infty)$, one can show that 
\begin{equation}
\label{deltaQ}
\delta Q := Q[\tilde{u},\tilde{v}] - Q[u, v] = - 32\,\kappa^7\,\frac{2^{9}}{7}.
\end{equation}
$\left(\mbox{making } \frac{7}{2^9\,\sqrt{2}}\,\delta Q = 
-\,2\,\left(4\,\kappa^{4}\right)^{7/4}\right)$. 
This result is similar to that for Schr\"{o}dinger operators
\cite{BenLoss} and reflects the loss of a doubly degenerate 
eigenvalue $E_{0} = -4\,\kappa^{4}$, when going from $L$ to
$\tilde{L}$.
The constant in the right hand side of \eqref{deltaQ} is related to 
the semiclassical constant
appearing in the trace formula for a fourth order differential operator 
considered in  \cite{Ost}.
\\
The proof of \eqref{deltaQ}, just as the derivation of
\eqref{functionalQ}, involves very lengthy calculations. When
deriving \eqref{deltaQ} one uses \eqref{odefg} and \eqref{darbouxL2}
to write the expression for $\delta Q$ as an integral of terms
involving only the functions $f$ and $g$, and their spatial derivatives. 
The crucial step is to note that the integrand is a pure derivative 
of $x$, i.e. $\delta Q = \int {\mathbb{Q}}\,' \, dx$ for some function 
$\mathbb{Q}$, which makes it possible to evaluate the integral solely 
from the limits of $f$ and $g$ at infinity. Thus, to compute $\delta
Q$, one selects the terms in $\mathbb{Q}$ which are free of
derivatives, as those are the only ones that contibute. The terms in 
$\mathbb{Q}$ still containing derivatives, for instance the ones
quadratic in $g$ and linear in $f$, 
\begin{alignat*}{4}
\int &\bigg((96-48)\,g^2\,f''' - 2\cdot96\,f\,g'\,g'' -
8\cdot12\,g''\,f'\cdot2g -4\cdot40\,f'''\,g^2 \\
&+ 160\,g''\,g'\,f - 
16\cdot26\,f''\,g'\,g - 16\cdot13\,g'^2\,f'\bigg)\,dx,
\end{alignat*}
give zero.

\section{Some examples.}
\textit{Example 1.} The operator
\begin{equation*}
L = \partial^4 - 5\,\partial^2 +
\partial\,\frac{12}{\cosh^2 x}\,\partial - \frac{6}{\cosh^2 x} =
A^{*} A - 4
\end{equation*}
with 
\begin{equation*}
A = - \partial^2 - 3\,\tanh x\,\partial - 2 
\end{equation*}
has 2 linearly independent eigenfunctions with eigenvalue $E_{0} = -4$, 
\begin{equation*}
\psi_{+}(x) = \frac{1}{\cosh^2 x}, \quad \psi_{-}(x) = 
\frac{\sinh x}{\cosh^2 x}.
\end{equation*}
One can easily check that $A\,\psi_{\pm} = 0$ and that $u, v$ are 
reflectionless, as 
\begin{equation*}
\tilde{L} = A A^{*} - 4 = \partial^4 - 5\,\partial^2
\end{equation*}
(note that $\psi_{+}$ and $\psi_{-}$ have different fall-off behaviour
at $\infty$ and that $W(x) = \cosh^{-3} x$). \\
\textit{Example 2.} The operator 
\begin{equation*}
L = \partial^4 + 16\,\partial\,\frac{1}{\cosh^2 x}\,\partial + 
\frac{40}{\cosh^4 x} - \frac{88}{\cosh^2 x} = A^{*} A - 64
\end{equation*}
with 
\begin{equation*}
A = - \partial^2 - 4\,\tanh x\,\partial - 8 + \frac{2}{\cosh^2 x}
\end{equation*}
has 2 linearly independent eigenfunctions with eigenvalue $E_{0} = -64$, 
\begin{equation*}
\psi_{+}(x) = \frac{\cos 2x}{\cosh^2 x}, \quad \psi_{-}(x) = 
\frac{\sin 2x}{\cosh^2 x}.
\end{equation*}
One easily verifies that $A\,\psi_{\pm} = 0$, and that 
\begin{equation*}
\tilde{L} = A A^{*} - 4 = \partial^4 - \frac{40}{\cosh^2 x}.
\end{equation*}
A computation gives that 
\begin{equation*}
Q = \frac{2^{10}}{7}\cdot 4 \cdot 687, \quad \tilde{Q} = 2^{10} \cdot
100, \quad \delta Q = -\frac{2^{21}}{7}  
\left(= - 32\,(\kappa = 2)^7\,\frac{2^9}{7}\right).
\end{equation*}\\
\textit{Example 3.} The operator
\begin{equation*}
L = \partial^4 + \left(45\,\Psi^4 - 40\,\Psi^2\right) = A^{*} A - 4
\end{equation*}
with 
\begin{equation*}
W = \Psi^{2} := \frac{1}{\cosh^{2} x}, \quad W_{12} = 2\,\Psi^2 - 3\,\Psi^4
\end{equation*}
and
\begin{equation*}
A = -\partial^2 - 2\,\tanh x\,\partial - 2  + 3\,\Psi^2 
\end{equation*}
has a doubly degenerate eigenvalue $E_{0} = -4$. One easily verifies,
that
\begin{equation*}
\tilde{L} = \partial^4 - 8\,\partial\,\Psi^2\,\partial +
25\,\Psi^4-16\,\Psi^2.
\end{equation*}
\textit{Example 4.} The operator
\begin{equation*}
L = \partial^4 - \partial^2 + 4\,\partial\,\frac{1}{\cosh^2
x}
\,\partial + \frac{6}{\cosh^{2} x} - \frac{8}{\cosh^{4} x} = 
A^{*} A 
\end{equation*}
with 
\begin{equation*}
A = -\partial^2 - \,\tanh x\,\partial - \frac{1}{\cosh^{2} x} =
\partial \left(-\partial - \tanh x\right)
\end{equation*}
has a unique ground-state $E_{0} = 0$ with eigenfunction 
\begin{equation*}
\psi (x) = \frac{1}{\cosh x}.
\end{equation*}
The second solution of $A\,\psi = 0$ is $\psi = \tanh x \not\in 
L^{2}\left(\mathbb{R}\right)$. One easily verifies, that 
\begin{equation*}
\tilde{L} = \partial^4 - \partial^2. 
\end{equation*}
\textit{Example 5.} For any $k > 0$, the operator
\begin{equation*}
L = \partial^4 + \partial\,u\,\partial + v 
\end{equation*}
with 
\begin{equation*}
\left\{
\begin{array}{rl}
u(x) &= 2\left(1 +
\frac{2}{k}\right)\,\Psi^{2}\left(\frac{x}{k}\right)\\
v(x) &= - 4\left(1 +\frac{1}{k} - \frac{1}{k^3}\right) 
\Psi^{2}\left(\frac{x}{k}\right) + \left(1 - \frac{1}{k}\right)\left(1
+ \frac{5}{k} + \frac{6}{k^2}\right)\Psi^{4}\left(\frac{x}{k}\right), 
\end{array}
\right.
\end{equation*}
where
\begin{equation*}
\Psi(x) := \frac{1}{\cosh x},
\end{equation*}
has a doubly degenerate ground-state, $E_{0} = -4$, with eigenfunctions
\begin{equation*}
\psi_{\pm}^{(k)}(x) = e^{\pm i x} \left(\frac{1}{\cosh \frac{x}{k}}\right)^{k}.
\end{equation*}
\section{Follytons.}
In order to find $u$ and $v$ such that $L = A^{*} A + E_{0}$ is
'conjugate' to the free operator $\tilde{L} = \partial^{4} =: L_{0}$
one has to solve \eqref{odefg} with $u = v = 0$. Eliminating $g$ and 
writing $E_{0} = - 4\kappa^{4}$ one obtains the ODE
\begin{equation*}
2\,f''' + 6\,f\,f'' + 7\,f'^2 + 8\,f'\,f^2 + f^4 = 16\,\kappa^{4}. 
\end{equation*}
One may reduce the order by taking $f$ as the independent variable,
and $F(f) := f'$ as the dependent one, yielding 
\begin{equation*}
2\left(F''\,F^2 + F'^2\,F\right) + 6\,F\,F'\,f + 7\,F^2 + 8\,F\,f^2 + f^4
= 16\,\kappa^4,
\end{equation*}
but both forms seem(ed) to be too difficult to solve. By using 
\eqref{hirota}, however, it takes the form
\begin{equation*}
16\kappa^4\,W^2 = 2\left(W''''\,W - W'''\,W'\right) + W''^2,
\end{equation*}
in which it is easier to see the solution \cite{Hoppe1}
\begin{equation*}
\hat{W} = \mbox{const } \cdot \left(\sqrt{2} + 
\cosh\left(2\,\kappa\,x\right)\right).
\end{equation*}
Correspondingly,
\begin{equation*}
\hat{f} := \frac{\hat{W}'}{\hat{W}} = 2\,\kappa \,
\frac{\sinh\left(2\,\kappa\,x\right)}{\sqrt{2} + 
\cosh\left(2\,\kappa\,x\right)}.
\end{equation*}
As interchanging $A^{*}$ and $A$ (as far as $f$ is concerned) only
changes the sign of $f$, 
\begin{equation*}
f (x) = -2\,\kappa \,
\frac{\sinh\left(2\,\kappa\,x\right)}{\sqrt{2} + 
\cosh\left(2\,\kappa\,x\right)}.
\end{equation*}
The Wronskian of the two ground-states $\psi_{\pm}$ (of $L =
\partial^4 + \partial\,u\,\partial + v$, conjugate to $L_{0} =
\partial^4$) is simply the inverse of $\hat{W}$, i.e. (choosing the 
constant in $\hat{W}$ to be $1$),
\begin{equation*}
W(x) = \frac{1}{\sqrt{2} + \cosh\left(2\,\kappa\,x\right)} =: 
\chi\left(2\,\kappa\,x\right).
\end{equation*}
The function $g$ is given by 
\begin{equation*}
g = \frac{1}{2} \left(3\,f' - f^2\right) = -2\,\kappa^2 \left(1 + 
\sqrt{2}\,W - 2\,W^2\right).
\end{equation*}
Insertion into equation \eqref{odefg} yields the reflectionless 
'potentials'
\begin{equation}
  \label{folly1}
\left\{
\begin{array}{rl}
u_{\kappa} &= 16\,\kappa^2 \left(\sqrt{2}\,W - W^2\right)\\
v_{\kappa} &= 16\,\kappa^4 \left(\sqrt{2}\,W - 12\,W^2 +
16\,\sqrt{2}\,W^3 - 8\,W^4\right)
\end{array}
\right.
\end{equation}
with $L = \partial^4 + \partial\,u_{\kappa}\,\partial + v_{\kappa}$
having exactly one doubly degenerate negative eigenvalue 
$-4\,\kappa^4$. While in most other examples we scaled $\kappa$ to be
equal to $1$ it is, in this case (due to the appearence of $2 \kappa$
in $W$) easiest to choose $\kappa = \frac{1}{2}$, i.e. to take 
\begin{equation}
  \label{folly2}
\left\{
\begin{array}{rl}
u &= 4 \left(\sqrt{2}\, \chi - \chi^{2}\right)\\  
v &= \left(\sqrt{2}\,\chi - 12\,\chi^2 + 16\,\sqrt{2}\,\chi^3-8\chi^4\right)
\end{array}
\right.
\end{equation}
and, when needed, use formulas like 
\begin{equation*}
\begin{array}{rl}
\chi '' &= \chi\,\left(1 - 3\,\sqrt{2}\,\chi + 2\,\chi^2\right)\\ 
\chi'^2 &= \chi^2\,\left(1 - 2\,\sqrt{2}\,\chi + \chi^2\right)\\
\chi''' &= \chi'\,\left(1 - 6\,\sqrt{2}\,\chi + 6\,\chi^2\right)\\
\chi'''' &= \chi\,\left(1 - 15\,\sqrt{2}\,\chi + 80\,\chi^2 - 
60\,\sqrt{2}\,\chi^{3} + 24\,\chi^4\right) .  
\end{array}
\end{equation*}
(Note that redefining $\chi$ by a factor of $-\sqrt{2}$ would make
all the coefficients positive (integers)). 
These formulas are useful when checking that $u(x + 4\,t)$ and $v(x + 4\,t)$,
with $u$ and $v$ given by \eqref{folly2}, are exact solutions of the 
non-linear system of PDE's \eqref{KdVtype} (just as 
$u_{\kappa}(x + 16\kappa^2 \,t), v_{\kappa}(x + 16\kappa^2\,t)$).

\appendix
\section{$W\not= 0$.}

We shall prove here that the Wronskian type function defined 
in \eqref{wronsk} never equals zero.

\medskip
\begin{theorem}\label{Wronsk} 
Let $\psi_\pm$ be two orthonormal eigenfunctions
of the operator \eqref{L} corresponding to the lowest eigenvalue 
$E_0$ of double multiplicity. 
Then 

\begin{equation}\label{wronsk1}
W[\psi_+,\psi_-](x):= \psi_{+}(x)\,\psi_{-}'(x) 
-  \psi_{-}(x)\,\psi_{+}'(x)  \not= 0, \quad x\in \mathbb{R}.
\end{equation}

\end{theorem}

\medskip
\noindent
In order to prove this result we need a simple auxiliary statement.
\medskip
\begin{lemma}\label{overdet}
Let $E_0$ be the lowest eigenvalue of the operator $L$ and 
let $\psi\in L^2(\mathbb{R})$ be a solution of the equation \eqref{groundstate}
satisfying $\psi(x_0) = \psi^{'}(x_0)= 0$ for some  $x_0\in \mathbb{R}$. 
Then $\psi(x) \equiv 0$.
\end{lemma}

\noindent
\textit{Proof}.
Indeed, the function
$$
\tilde \psi(x) =
\begin{cases}
\psi(x), & \text{if \(x\leq x_0\),} \\
-\psi(x), & \text{if \(x\geq x_0\),} 
\end{cases}
$$
is linear independent with $\psi$. Since 
$\tilde\psi(x_0) = \tilde\psi^{'}(x_0)= 0$ we obtain
$(L \tilde\psi, \tilde\psi)  = E_0\|\tilde\psi\|^2$.
Then $\tilde\psi$ is also an eigenfunction of the operator $L$ with 
the eigenvalue $E_0$. Consider now the linear combination
$$
\psi_1(x) = \tilde\psi^{''}(x_0)\psi(x) - \psi^{''}(x_0)\tilde\psi(x).
$$
Obviously $\psi_1(x_0)=\psi_1{'}(x_0)=\psi_1{''}(x_0)=0$, $\psi_1\in 
L^2\left(\mathbb{R}\right)$
and $\psi_1$  satisfies the fourth order differential equation $L\psi_1 = E_0\psi_1$.
Being overdetermined, $\psi_1\equiv 0$ which also implies $\psi\equiv 0$. $\Box$

\medskip
\noindent
{\it Remark}.
In Lemma~\ref{overdet} the conditions $\psi(x_0) = \psi^{'}(x_0)= 0$
split the problem for the operator $L$ 
in $L^2(\mathbb{R})$ into two Dirichlet boundary value problems on semiaxes
$L^2((x_0,\infty))$ and $L^2((-\infty, x_0))$. Therefore,  
the lowest eigenvalue moves up.

\medskip
\noindent
{\it Proof of Theorem~\ref{Wronsk}}.

\noindent
{\it a.} Let $\psi_\pm$, be two orthonormal eigenfunctions 
corresponding to the lowest eigenvalue $E_0$ of the operator $L$.
The functions $\psi_+$ and $\psi_-$ cannot vanish at the same point. Indeed,
assume that they do. Then there is a point $x_0$ such 
that $\psi_+(x_0)=\psi_-(x_0) = 0$.
If in addition we assume that say $\psi_+^{'}(x_0)=0$, 
then by Lemma~\ref{overdet} $\psi_+\equiv 0$ and we obtain a contradiction.
Therefore we can assume that $\psi_\pm^{'}(x_0) \not=0$. Introduce a new 
function 
$$
\psi_2(x) = \psi_-^{'}(x_0) \psi_+(x) - \psi_+^{'}(x_0)\psi_-(x).
$$
It is a non-trivial eigenfunction of the equation \eqref{groundstate} satisfying
$\psi_2(x_0)=\psi_2^{'}(x_0)=0$. By using Lemma~\ref{overdet} again 
we find that $\psi_2\equiv0$ which cannot be true because  $\psi_+$ and $\psi_-$
are linear independent. 

\smallskip
\noindent
{\it b.} Consider now the following pair of complex functions
$$
\Psi_{\pm}(x) = \psi_+(x) \pm i\psi_-(x) =: \psi(x) e^{\pm i \phi(x)}.
$$
By using {\it a}. we observe that $\psi$ never 
vanishes, $\psi(x)\not= 0$, $x\in \mathbb{R}$. Besides
$$
W[\Psi_+, \Psi_-] =
(\psi_+ +i\psi_-) (\psi_+ -i\psi_-)^{'} - (\psi_+ +i\psi_-)^{'} (\psi_+ -i\psi_-) 
$$
$$
= -2i \,W[\psi_+,\psi_-]  = -2i \,\phi^{'}\psi^2.
$$
Thus, in order to prove Theorem \ref{Wronsk} it remains to prove that 
$\phi^{'}\not=0$.
Assume that there is $x_0$ such that $\phi^{'}(x_0) = 0$ and consider
$$
\Phi(x) = e^{-i\phi(x_0)}\Psi_+(x) - e^{i\phi(x_0)}\Psi_-(x). 
$$
Clearly $\Phi(x_0)=\Phi^{'}(x_0)=0$ and by using Lemma~\ref{overdet} we 
obtain $\Phi\equiv0$ which contradicts the linear independency of the 
functions $\Psi_\pm$.

\noindent
The proof is complete. $\Box$

\medskip
\noindent
{\it Acknowledgments.}
The authors would like to thank H. Kalf, E. Langmann, A. Pushnitski and 
O. Safronov for useful discussions, as well as the ESF European
programme SPECT and the EU Network: ``Analysis \& Quantum" for partial 
support.

\end{document}